\documentclass[a4paper]{article}
\usepackage{graphicx,cite,amsmath}
\usepackage[margin=2.5cm]{geometry}
\usepackage[utf8]{inputenc}
\graphicspath{{figs/}}

\title{Exploring the QCD phase diagram via the collision energy dependence of
multi-particle femtoscopy with PHENIX}

\author{M\'at\'e Csan\'ad for the PHENIX Collaboration\\
E{\"o}tv{\"o}s Lor{\'a}nd University, H-1117 Budapest, P{\'a}zm{\'a}ny P. s. 1/A, Hungary}

\begin{document}
\maketitle

\begin{abstract}
Exploration of the rich structure of the QCD phase diagram is an important topic in the RHIC heavy ion program. One of the ultimate goals of this program is to search for the critical endpoint. Investigation of the space-time structure of hadron emissions at various phase transition points using Bose-Einstein correlations of identical bosons may provide insight on the location of the critical endpoint. PHENIX has performed comprehensive measurements of the Bose-Einstein correlation in Au+Au collisions at $\sqrt{s_{NN}}$ = 15, 19, 27, 39, 62.4, and 200 GeV, where we incorporated Lévy-type source functions to describe the measured correlation functions. We put particular focus on one of the parameters of the Lévy-type source functions, the index of stability $\alpha$, which is related to one of the critical exponents (the so-called correlation exponent $\eta$). We have measured its collision energy and centrality dependence. We have also extended our analysis from two-particle to three-particle correlations to characterize the nature of the hadron emission source. The three particle correlations confirmed the findings of the two-particle correlations, and also provide insight on the pion production mechanism beyond the core-halo model.  
\end{abstract}

\section{Introduction}

Femtoscopy (coined by Ledniczky~\cite{Lednicky:2001qv}) is an important subfield of high energy nuclear and particle physics, as it allows us to investigate the space-time structure of femtometer scale processes. This subfield originates from the work of Hanbury Brown and Twiss, who investigated the angular diameter of stars using radio and optical telescopes, based on intensity correlations; this work was later on understood by Glauber, and the technique independently discovered by Goldhaber and collaborators (see more details and references  in Ref.~\cite{Csanad:2018vgk}). Nowadays we understand femtoscopic correlations to be caused by Bose-Einstein statistics, and hence use the name Bose-Einstein or quantum statistical correlations as well.

The most important equation utilized in two-particle femtoscopic correlations is one that relates the pair source $D_2(r,K)$ (describing the probability density of creating a pair with spatial separation of $r$ and average momentum $K$) and the correlation function $C_2(q,K)$ (indicating the amount of correlation of pairs with momentum difference $q$ and average momentum $K$):
\begin{align}
C_2(q,K) = 1 + \frac{\widetilde{D_2}(q,K)}{\widetilde{D_2}(0,K)},\label{e:C2D}
\end{align}
where $\widetilde{D_2}$ denotes the Fourier-transform of $D_2$ in its first variable. The $C_2$ correlation function can also be written up with the single particle source $S(r,p)$ ($D_2(r,K)$ is the autoconvolution of $S(r,p)$ in the first variable with $p=K$):
\begin{align}
C_2(q,K) = 1 + \left|\frac{\widetilde{S}(q,K)}{\widetilde{S}(0,K)}\right|^2.\label{e:C2S}
\end{align}
There are several approximations behind these equations, see more details and references e.g. in section II.D of Ref.~\cite{Adare:2017vig}: some of these are related to the kinematics of the pair (e.g. requiring the particle momenta to be close to each other), others to thermal emission. Given the validity of these assumptions, the above equations indicate the importance of femtoscopy: by measuring $C_2$, one can obtain information on femtometer-scale processes via the above mentioned $S$ or $D_2$ source functions. If one assumes a Gaussian source for example, then the correlation function will also be Gaussian, and hence the Gaussian source radius can be directly measured. Generally, for expanding sources the obtained radius will be related to the homogeneity length of the source for particles of given momenta~\cite{Adare:2017vig}.

\begin{figure}
\vspace{-10pt}
\begin{center}
\includegraphics[width=0.67\linewidth]{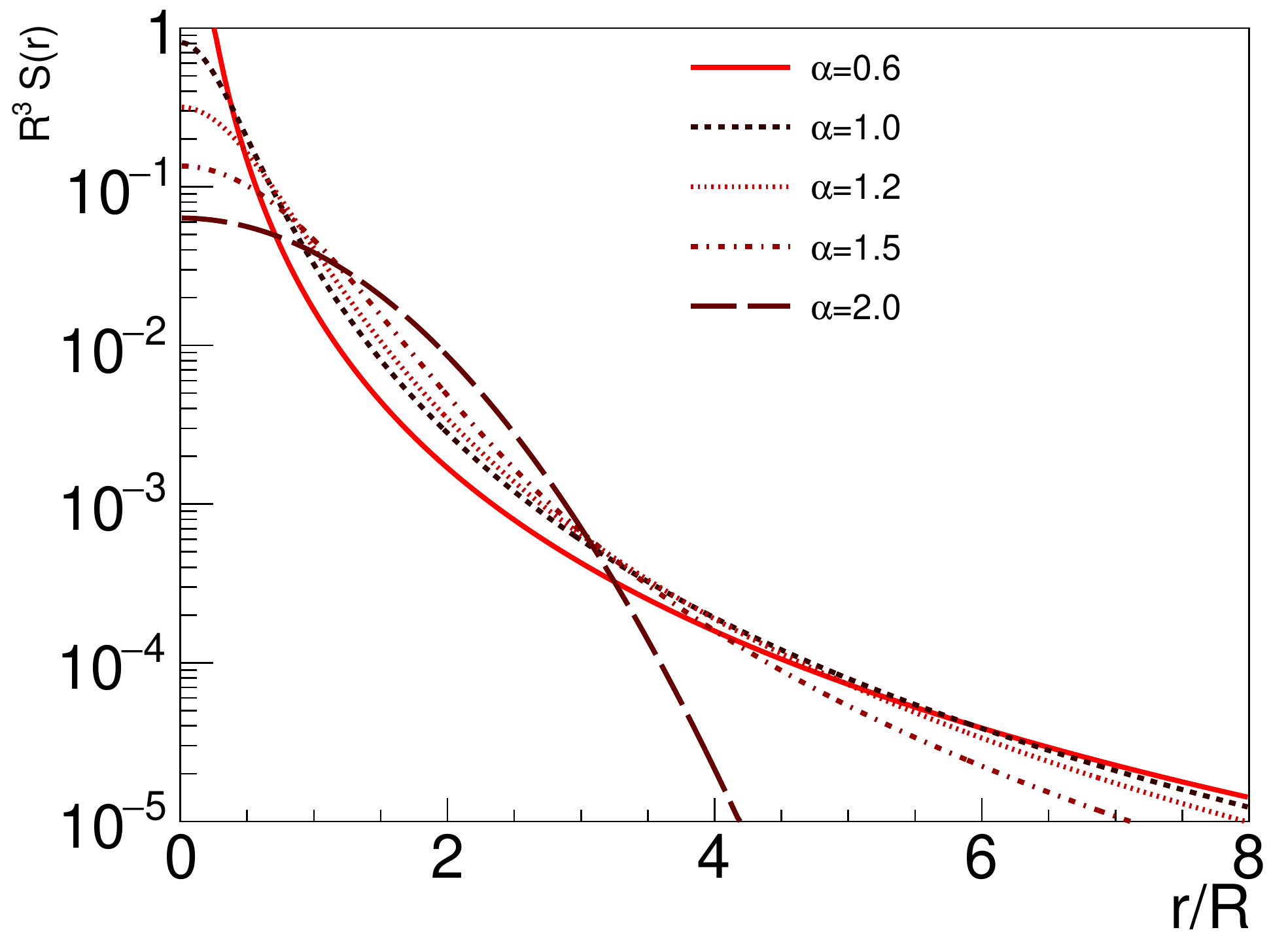}
\end{center}
\vspace{-15pt}
\caption{Spherically symmetric Lévy sources $\mathcal{L}(\alpha,R,r)$ for various $\alpha=$ values. The variable on the horizontal axis is $r/R$, since the $R$ dependence can be scaled out that way.}
\label{f:sourceplot}
\end{figure}

It was observed~\cite{Adler:2006as,Afanasiev:2007kk,Achard:2011zza,Sirunyan:2017ies,Adare:2017vig} that when investigating two-particle correlation functions, one has to go beyond the Gaussian approximation. Lévy-distributed sources with power-law tails may appear (among other phenomena) due to anomalous diffusion~\cite{Metzler:1999zz,Csorgo:2003uv,Csanad:2007fr}. These are characterized by the L\'evy index $\alpha$, and include Gaussian ($\alpha=2$) as well as Cauchy ($\alpha=1$) distributions as well. The three-dimensional L\'evy source is given as
\begin{align}\label{e:Levydef}
\mathcal{L}(\alpha,R,\mathbf{r})=\frac{1}{(2\pi)^3}\int \mathrm{d}^3\mathbf{q}\, e^{i\mathbf{q}\mathbf{r}} e^{-\frac{1}{2}|\mathbf{q} \mathbf{R^2} \mathbf{q}|^{\alpha/2}},
\end{align}
where $\mathbf{R^2}$ is the matrix of homogeneity lengths (also known as femtoscopic radii), which (in case of spherical symmetry) can be assumed to be the product of a single L\'evy scale squared ($R^2$) and the $3\times 3$ identity matrix, and then the last exponent will contain $(|\mathbf{q}| R)^\alpha$. In this case $\mathcal{L}(\alpha,R,\mathbf{r})$ becomes spherically symmetric. Such spherically symmetric L\'evy sources with various $\alpha$ values are illustrated in Fig.~\ref{f:sourceplot}. If one aims to describe more general sources, a convenient assumption is to introduce the radii in the Bertsch-Pratt coordinate system $R_{\rm out}$, $R_{\rm side}$ and $R_{\rm long}$, which represent the square roots of the diagonal elements of the $\mathbf{R^2}$ matrix. A 3D femtoscopy analysis usually uses this definition. One may go even beyond this approximation and introduce off-diagonal elements, but we refrain from doing that in the analyses presented in this paper.

Besides anomalous diffusion, other effects can also cause the appearance of Lévy sources. One important possibility is the QCD critical point, where large scale spatial fluctuations can be present, and the spatial correlation function becomes a power-law~\cite{Csorgo:2005it,Csorgo:2009gb}. In this case the $\eta$ critical exponent may be identical to the Lévy exponent $\alpha$. Finite size effects or dynamical criticality may distort this simple picture; nevertheless, this connection between $\eta$ and $\alpha$ yields a strong motivation to measure the Lévy index precisely. For more details, see Section IV.B of Ref.~\cite{Adare:2017vig}.

As seen in Eqs.~\eqref{e:C2D} and \eqref{e:C2S}, the momentum correlation function takes the value 2 for $q=0$. Experimentally however, one cannot measure correlations down to $q=0$ due to momentum resolution effects. The extrapolated value of $C_2(q\rightarrow 0)$ differs generally from 2, and takes the value $1+\lambda$. This can be interpreted in the core-halo model easily: it appears that the quantity $\lambda=C_2(q\rightarrow 0)-1$ is related to the ratio of directly produced hadrons (of the given species used in the measurement) to decay hadrons (of the same species) from long lived resonances. The former particles form the core, while the latter form the halo, and in fact $\lambda=f_c^2$, where $f_c=\textrm{core}/(\textrm{core+halo})$. This relation makes it possible to study the in-medium mass modification of the $\eta'$ meson, as detail in Section IV.A of Ref.~\cite{Adare:2017vig}.

Furthermore, it is important to note that the above noted $\lambda$ parameter may differ from unity due to other reasons as well. One such possibility is coherent pion production~\cite{Bolz:1992hc,Weiner:1999th,Csorgo:1999sj}. In this case the above mentioned approximations are not valid, and hence the strength of multi-particle Bose-Einstein correlations ($C_n$) will be modified; these correlation strengths are given by
\begin{align}
\lambda_n = \lim_{q_{ij}\rightarrow 0}C_n(\{q_{ij}\})-1,
\end{align}
where $\lambda_2\equiv\lambda$ for simplicity. In a higher order correlation function, lower order correlations will also give contributions, for example pairs within triplets or quadruplets. Hence a given $\lambda_n$ has contributions from lower order correlations as well. If $p_c$ is defined as the fraction of coherently produced pions (within the core), then for 2- and 3-particle correlations one obtains~\cite{Csorgo:1999sj}
\begin{align}
\lambda_2 &= f_c^2\left[(1-p_c)^2+2p_c(1-p_c)\right] = f_c^2(1-p_c^2),\label{e:lambda2fcpc}\\
\lambda_3 &= 2f_c^3\left[(1-p_c)^3+3p_c(1-p_c)^2\right] + 3f_c^2\left[(1-p_c)^2+2p_c(1-p_c)\right].
\end{align}
This means that simultaneous measurement of 2- and 3-particle correlations may shed light on the value of $p_c$ and $f_c$~\cite{Csanad:2018vgk}. Furthermore, one may define a core-halo independent parameter as
\begin{align}
\kappa_3 = \frac{\lambda_3 - 3\lambda_2}{2\sqrt{\lambda_2}^3} \label{e:kappa3}
\end{align}
which is independent of the value of $f_c$, and it takes the value $\kappa_3=1$ if $p_c=0$. Hence by measuring $\kappa_3$, one can investigate if the simple core-halo model interpretation of $\lambda_2$ is consistent.

\section{Results}
Two-pion measurements were performed with the Au+Au data taken by the PHENIX detector system during the 2010-2011 running period. Correlation functions were measured using the event mixing technique, and fits based on Lévy-distributed sources were done to obtain the statistically most probable Lévy parameters $\alpha$, $R$ and $\lambda$. Details about the analyses presented in this paper are given in Refs.~\cite{Adare:2017vig,Kincses:2017zlb,Kincses:2018vuo,Kurgyis:2018zck,Csanad:2018vgk,Lokos:2018dqq,Kurgyis:2019xzt}.

\begin{figure}
\begin{center}
\includegraphics[width=0.495\linewidth]{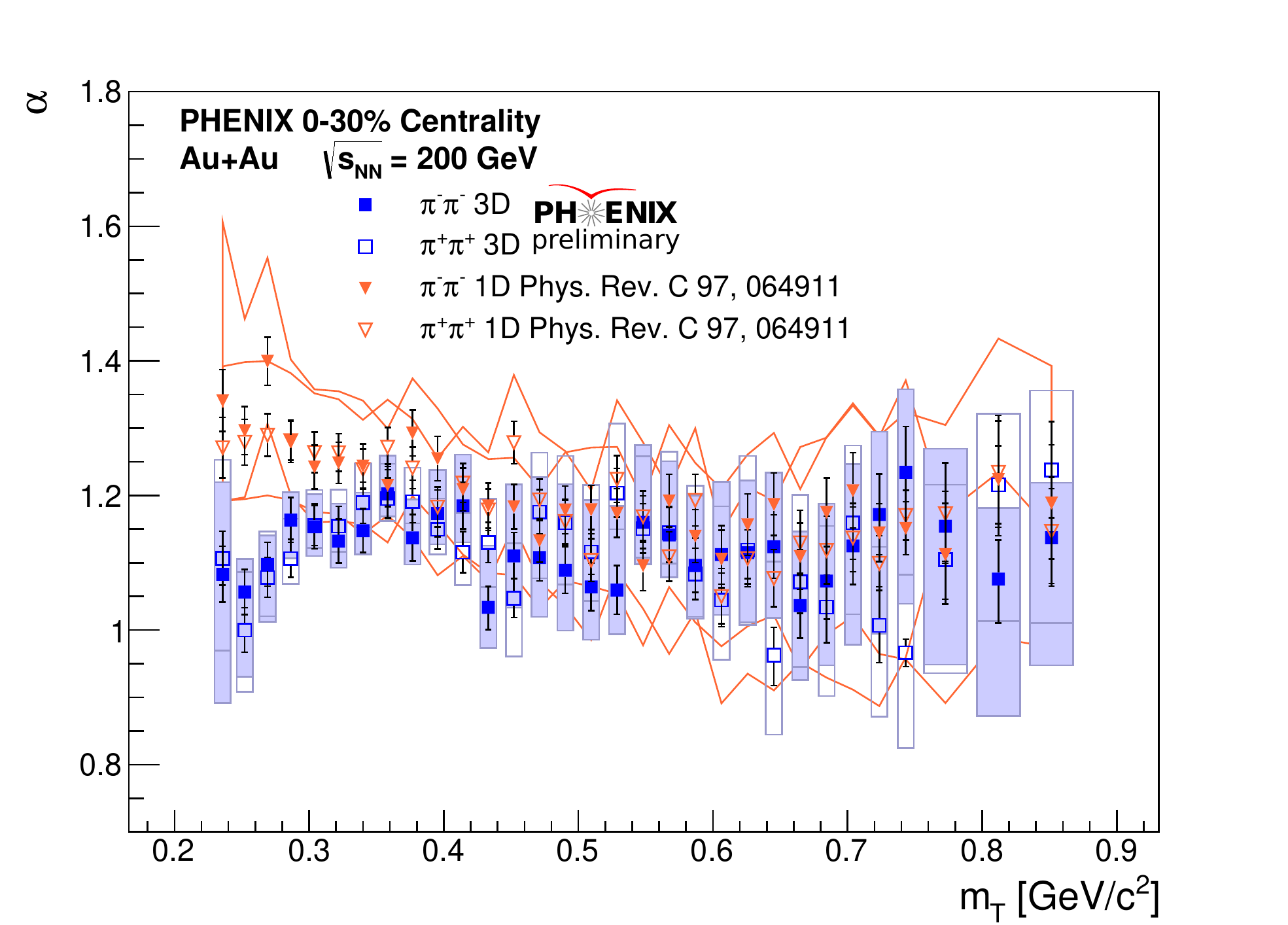}
\includegraphics[width=0.495\linewidth]{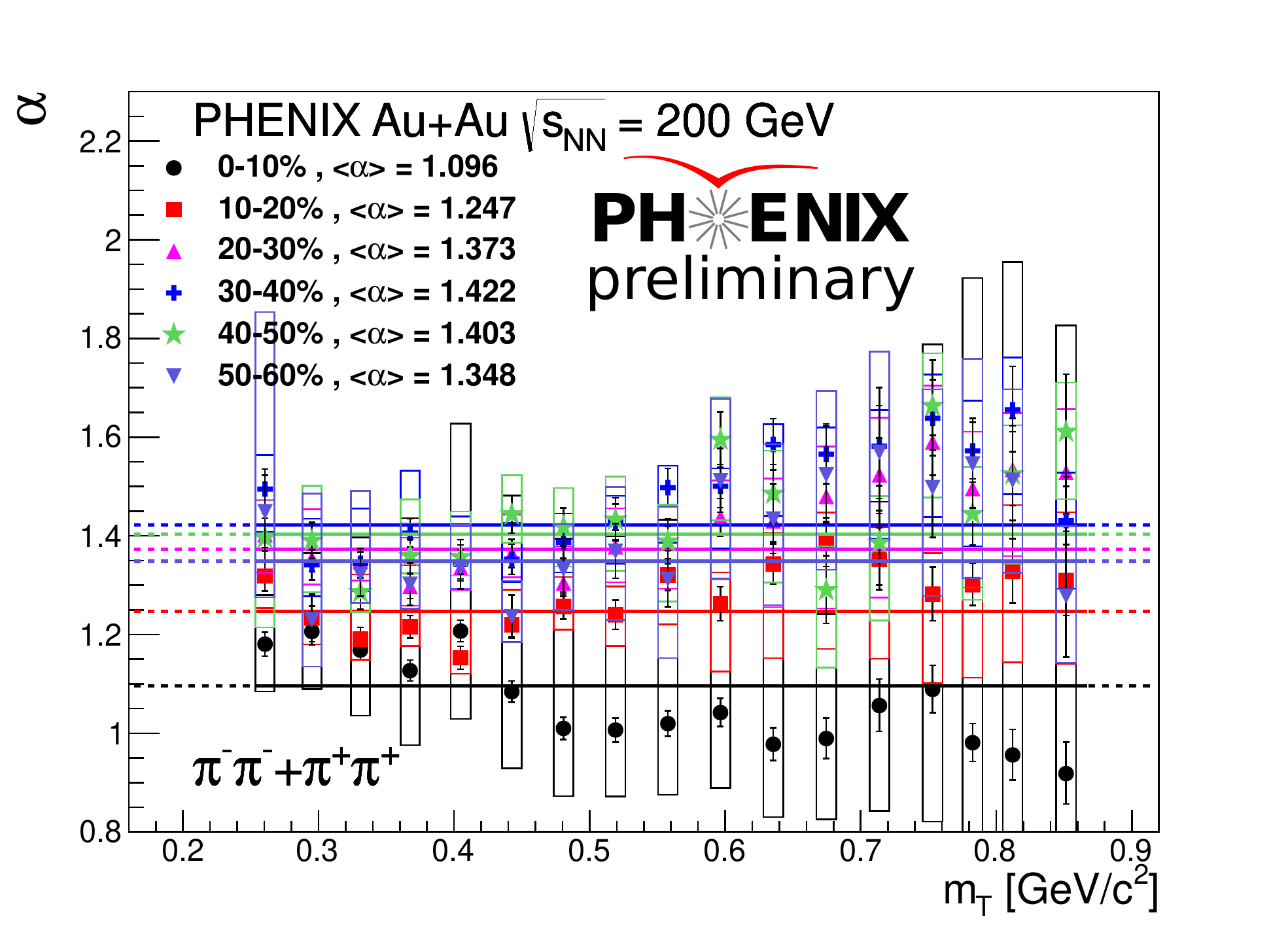}
\includegraphics[width=0.495\linewidth]{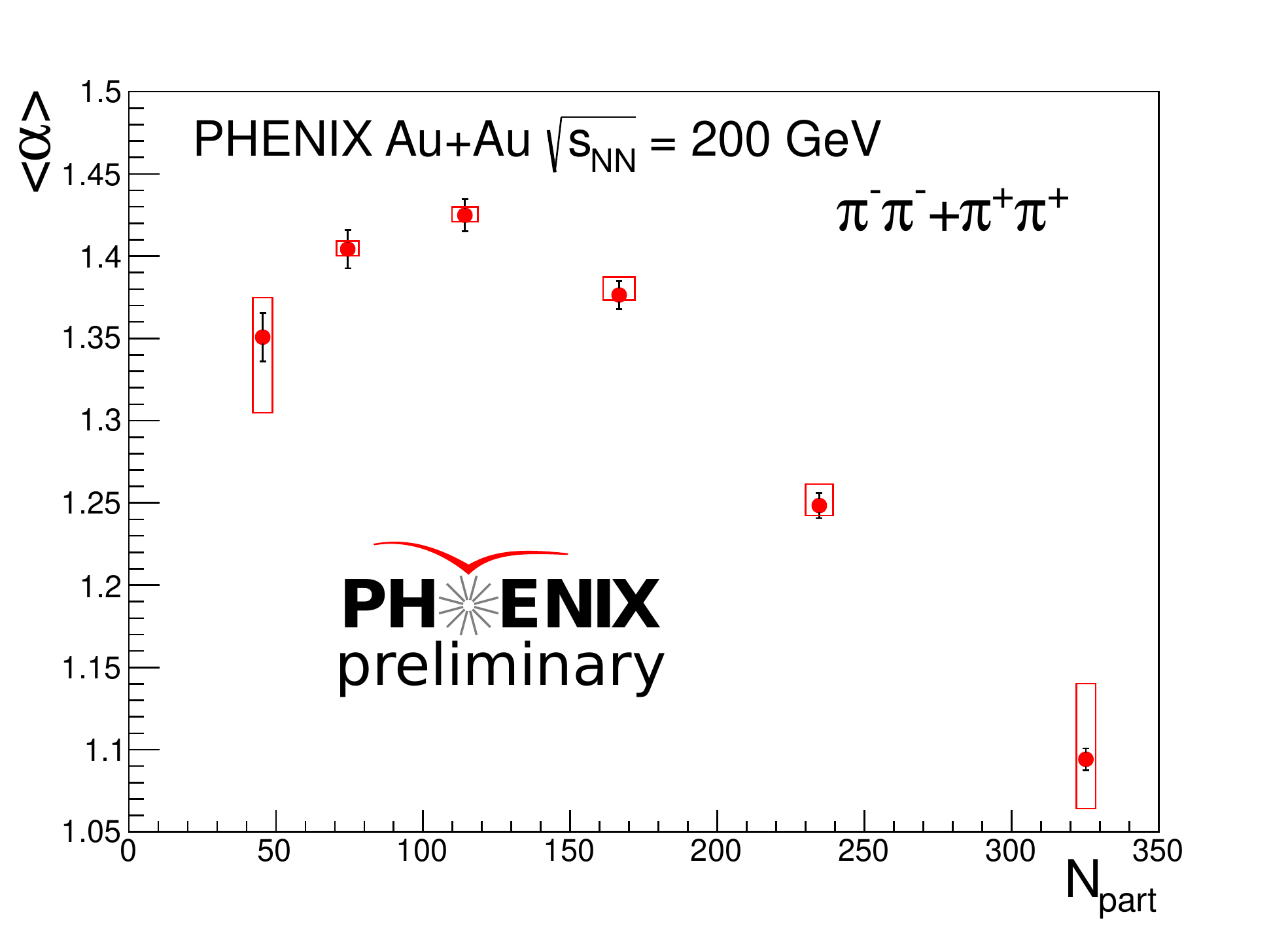}
\includegraphics[width=0.495\linewidth]{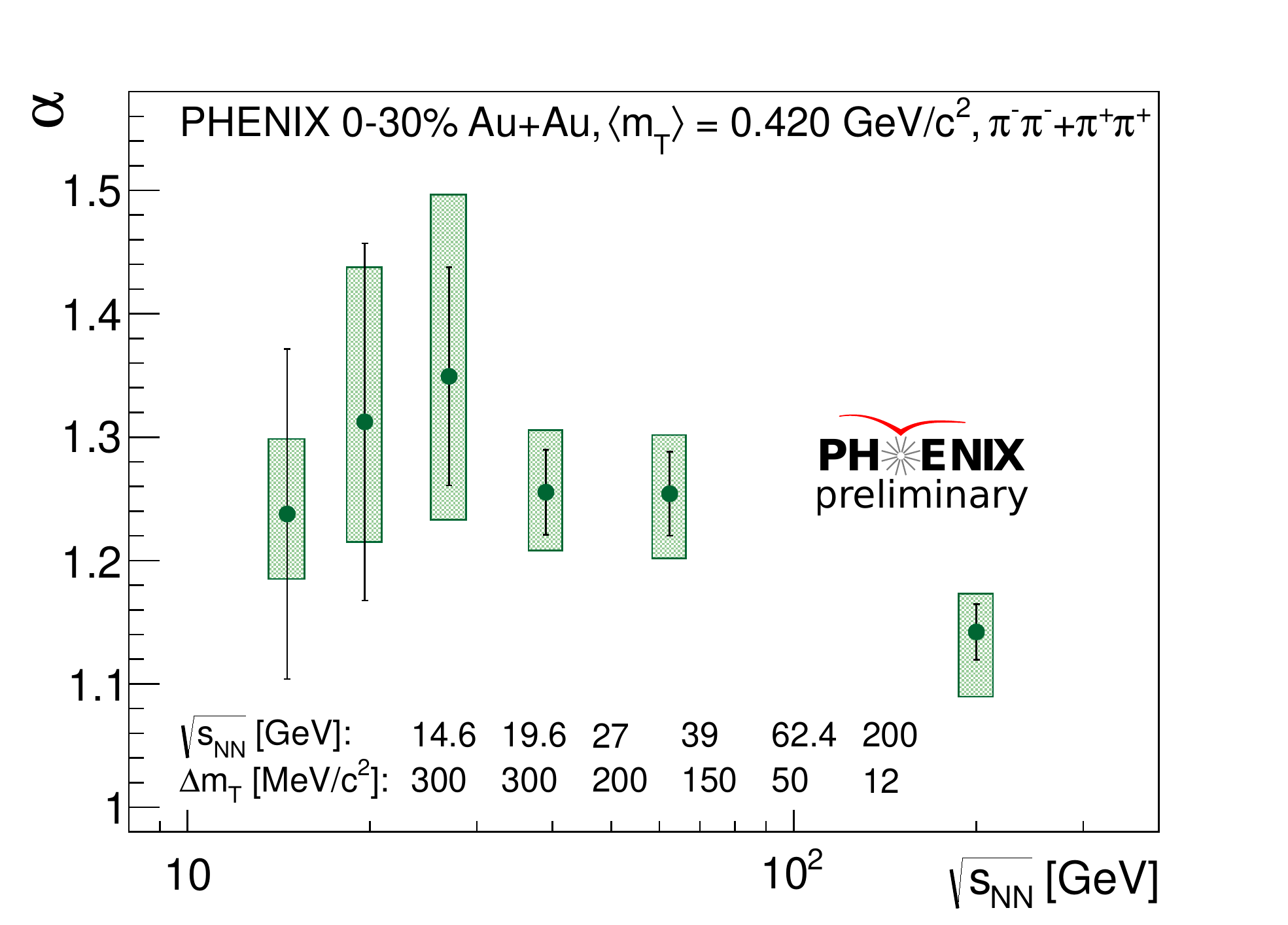}
\end{center}
\vspace{-10pt}
\caption{Measurements of the Lévy exponent $\alpha$ in 0-30\% centrality $\sqrt{s_{_{NN}}}=200$ GeV Au+Au collisions (top left panel), various 10\% centrality bins also at 200 GeV (versus $m_T$ in the top right panel, versus $N_{\rm part}$ in the bottom left panel), and in 0-30\% centrality collisions in for various $\sqrt{s_{_{NN}}}$ values (bottom right panel).}
\label{f:alpha}
\end{figure}


\begin{figure}
\begin{center}
\includegraphics[width=0.99\linewidth]{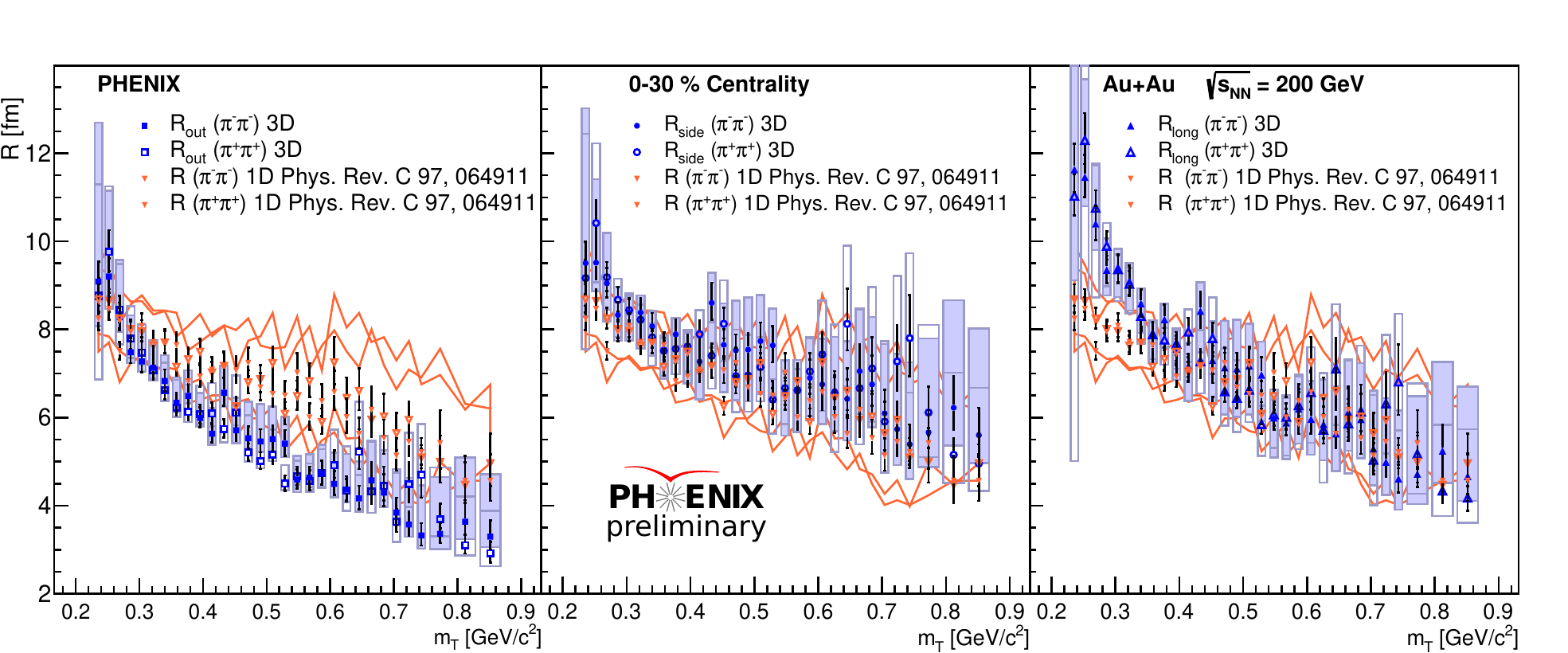}
\end{center}
\vspace{-10pt}
\caption{Transverse momentum dependence of the Bose-Einstein correlation radii (Lévy scale parameters) $R_{\rm out, side, long}$ in $\sqrt{s_{_{NN}}}=200$ GeV Au+Au collisions. Results of the 1D analysis (with a spherically symmetric source, employing only one Lévy scale $R$) are also shown in each of the panels for comparison.}
\label{f:r}
\end{figure}

\begin{figure}
\begin{center}
\includegraphics[width=0.49\linewidth]{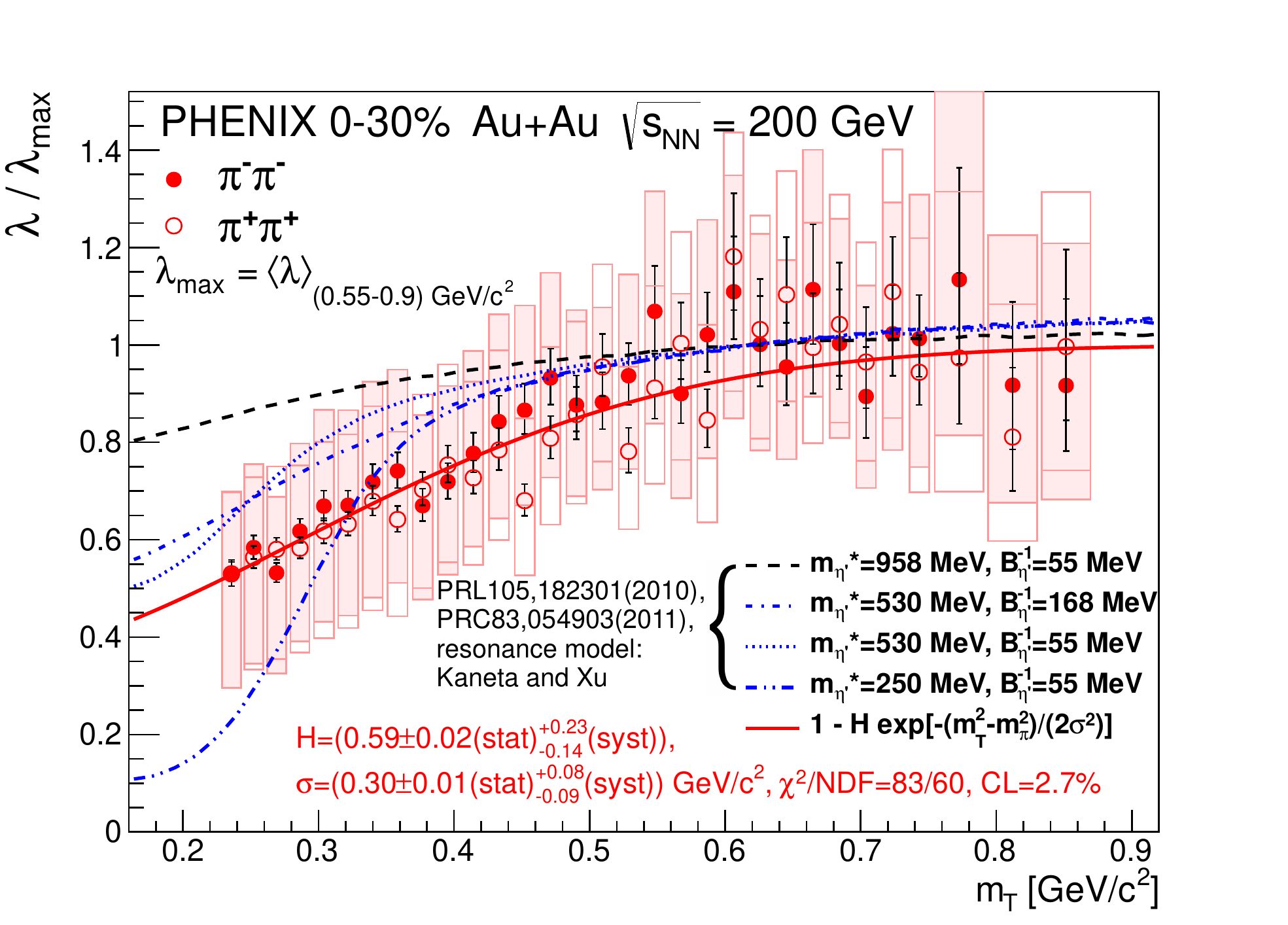}
\includegraphics[width=0.49\linewidth]{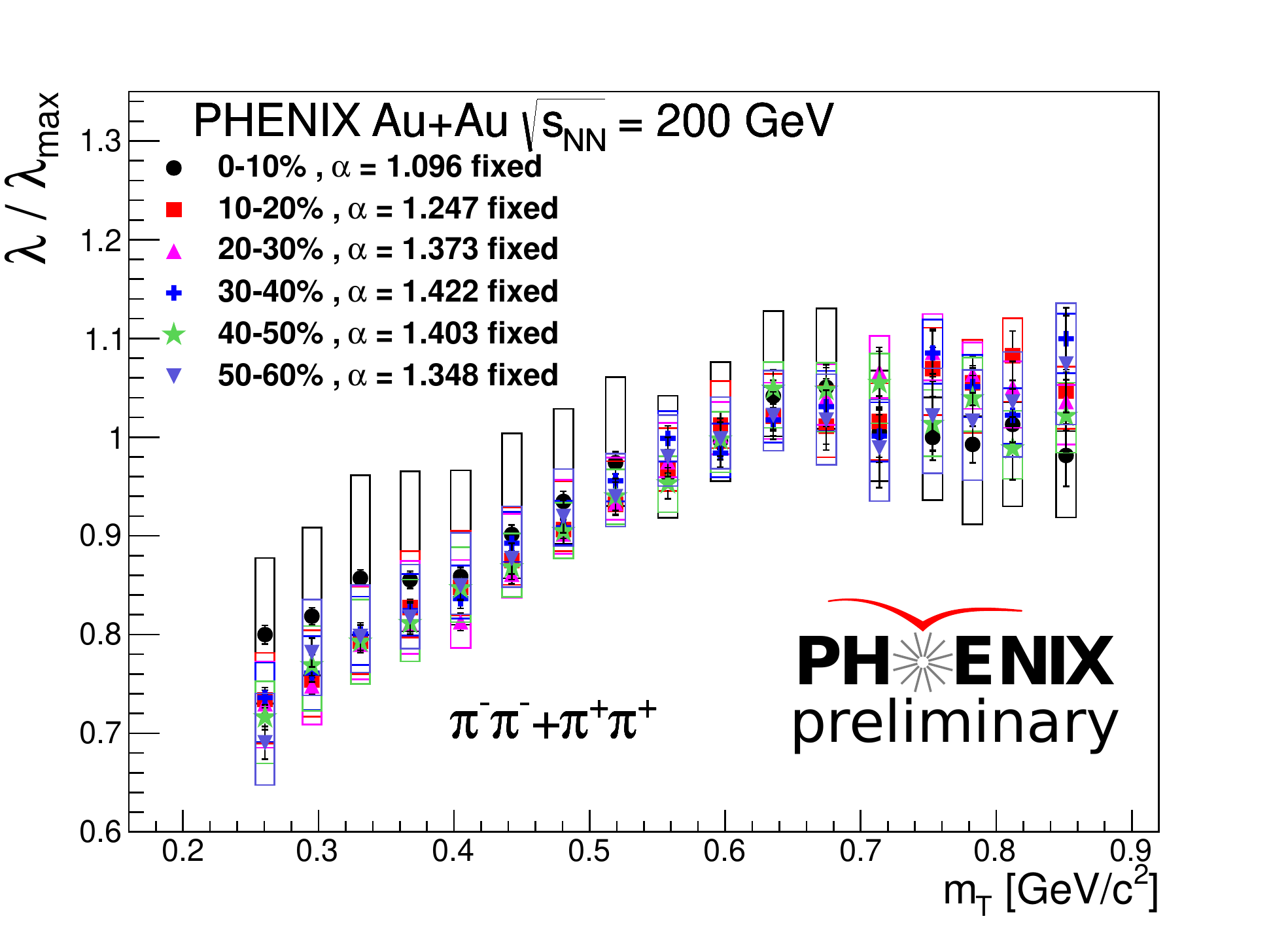}\\
\includegraphics[width=0.49\linewidth]{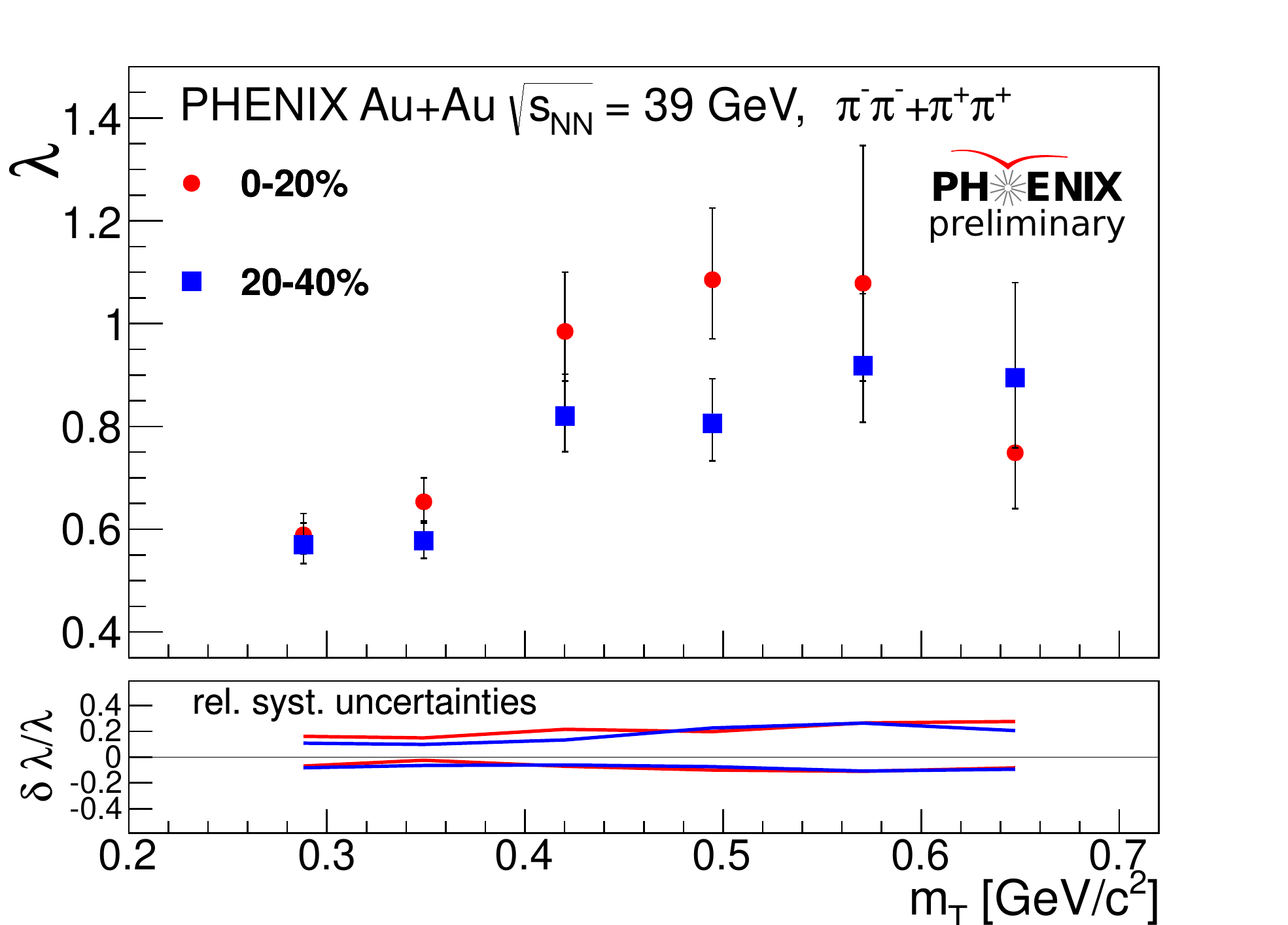}
\includegraphics[width=0.49\linewidth]{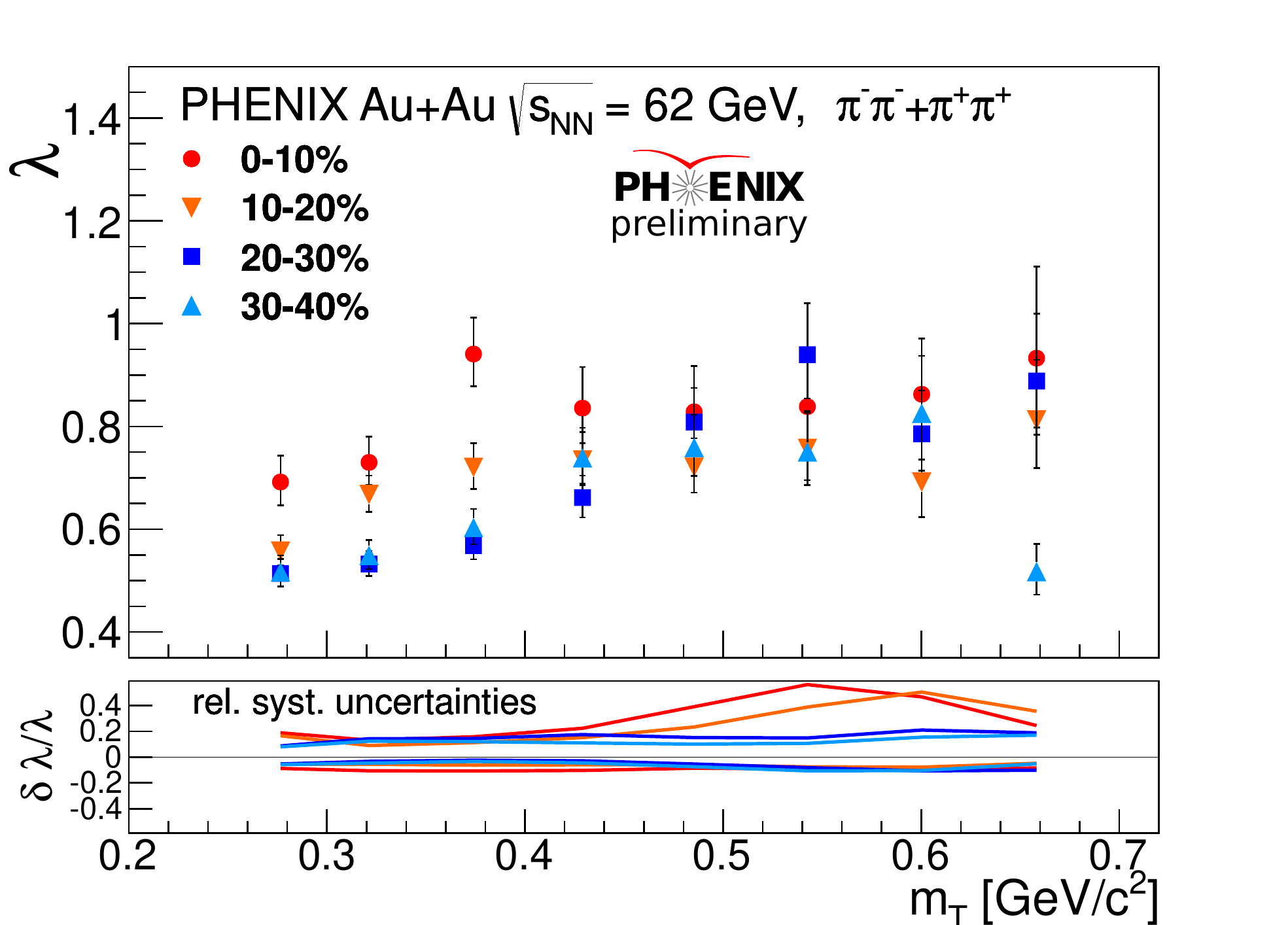}\\
\vspace{25pt}
\includegraphics[width=0.75\linewidth]{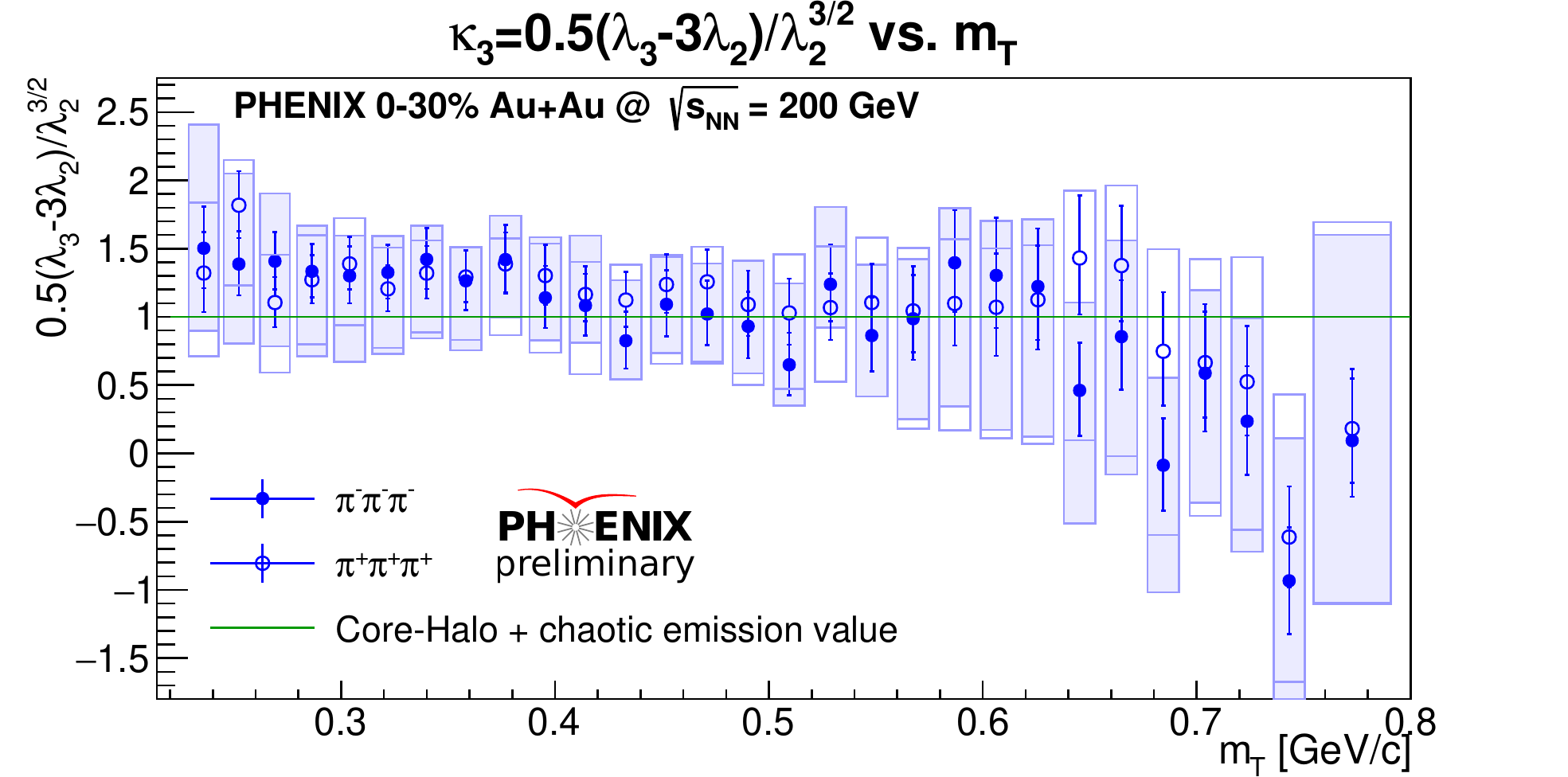}
\end{center}
\caption{Results on $\lambda$ normalized by its large $m_T$ average ($\lambda_{\rm max}$) in $\sqrt{s_{_{NN}}}=200$ GeV Au+Au collisions are shown in the top two plots. The top left plot also shows calculations based on various $\eta'$ masses. The second row shows $\lambda(m_T)$ for various centralities in $\sqrt{s_{_{NN}}}=39$ and 62 GeV Au+Au collisions. The bottom plot shows the core-halo independent $\kappa_3$ parameter as a function of $m_T$ (here for clarity, $\lambda_2\equiv\lambda$ is the strength of two-particle correlations, while $\lambda_3$ is the strength of three-particle correlations).}
\label{f:lambda}
\end{figure}

The Lévy exponent $\alpha$ was measured in collisions of various center of mass energies ($\sqrt{s_{_{NN}}}$ ranging from 14.5 GeV to 200 GeV) and with various centrality selections, as shown in Fig.~\ref{f:alpha}. These plots show that:
\begin{itemize}
  \setlength{\parskip}{0pt}
  \setlength{\parsep}{0pt}
  \setlength{\itemsep}{0pt}
  \vspace{-3pt}
\item Results on $\alpha$ from the 1D and the 3D analyses are compatible with each other, confirming the validity of the applied approach.
\item While $\alpha(m_T)$ is approximately constant for each centrality, there is a clear centrality dependence, with  $\alpha$ decreasing for more central collisions.
\item There does not seem to be a strong $\sqrt{s_{_{NN}}}$ dependence in this range - but this conclusion can be affected by the different $m_T$ windows used for each of the analysis (this had to be done as there was not enough statistics at the lower collision energies). 
\end{itemize}
  \vspace{-3pt}

The Lévy scale $R$ was also measured in both 1D and 3D analyses, and the results are shown in Fig.~\ref{f:r}. The 1D (spherically symmetric) analysis is mostly compatible with the 3D results, but there are clear differences in $R_{\rm out}$ in the $m_T\in [0.3-0.7]$ GeV$/c^2$ range, as well as for $R_{\rm long}$ for the $m_T<0.35$ GeV$/c^2$ range. This may be due to the non-spherical nature of the source in the longitudinally comoving system (LCMS). This may also be the reason for the small discrepancy between the 1D and the 3D results on $\alpha(m_T)$ seen in Fig.~\ref{f:alpha}.

The correlation strength parameter $\lambda$ was measured versus $m_T$ from $\sqrt{s_{_{NN}}}=39$ GeV to 200 GeV, and the results are shown in Fig.~\ref{f:lambda}. The results show a reduced $\lambda$ and small $m_T$ values, as compared to larger $m_T$ values; this seems to be an universal feature of $\lambda(m_T)$ across collision energies and centralities in the investigated range. As discussed in Ref.~\cite{Adare:2017vig}, this may be attributed to an in-medium mass reduction of the $\eta'$ meson. This statement relies heavily on the core-halo model assumption. This was further tested in a simultaneous analysis of 2- and 3-particle correlations, in particular the measurement of the $\kappa_3$ parameter defined above in Eq.~\eqref{e:kappa3}. The lowest panel of Fig.~\ref{f:lambda} shows that this parameter is compatible with $\kappa_3=1$, i.e. the value predicted by the core-halo model with fully chaotic emission.

\section{Acknowledgments}
The speaker was supported  by the ÚNKP-19-4 New National Excellence Program of the Hungarian Ministry for Innovation and Technology, the János Bolyai Research Scholarship of the Hungarian Academy of Sciences, as well as grant FK-123842 of the National Research, Development and Innovation Office of Hungary.

\end{document}